\documentclass[pre,aps,onecolumn,superscriptaddress,nofootinbib]{revtex4}%{revtex4-1}
\pdfoutput=1
\usepackage{amssymb}
\usepackage{epsfig}
\usepackage{amsmath}
\usepackage{subfigure}
\usepackage{graphicx}
\usepackage{textcomp}
\usepackage{url}
\usepackage{float}
\usepackage{color}
\usepackage{cases}
\usepackage[normalem]{ulem}

\usepackage[titletoc,title]{appendix}
\usepackage[colorlinks=true, urlcolor=blue, anchorcolor=blue, citecolor=blue,filecolor=blue,linkcolor=blue,menucolor=blue%,pagecolor=blue
]{hyperref}

\usepackage{color}
\begin{document}
\title{Mortal Brownian motion: three short stories}

\author{Baruch Meerson}
\email{meerson@mail.huji.ac.il}
\affiliation{Racah Institute of Physics, Hebrew University of
Jerusalem, Jerusalem 91904, Israel}

%\pacs{05.40.-a, 05.70.Np, 68.35.Ct}

\begin{abstract}
Mortality introduces an intrinsic time scale into the scale-invariant Brownian motion. This fact has important consequences for different statistics of Brownian motion. Here we are telling three short stories, where spontaneous death, such as radioactive decay, puts a natural limit to ``lifetime achievements" of a Brownian particle. In story 1 we determine the probability distribution of a mortal Brownian particle (MBP) reaching a specified point in space at the time of its death. In story 2 we determine the probability distribution of the area  $A=\int_0^{T} x(t) dt$ of a MBP on the line.  Story 3 addresses the distribution of the winding angle of a MBP wandering around a reflecting disk in the plane. In stories 1 and 2 the probability distributions exhibit integrable singularities at zero values of the position and the area, respectively. In story 3 a singularity at zero winding angle appears only in the limit of very high mortality. A different integrable singularity appears at a nonzero winding angle. It is inherited from the recently uncovered singularity of the short-time large-deviation function of the winding angle for immortal Brownian motion.

\end{abstract}

\maketitle
\nopagebreak

\section{Introduction}
In the recent years there has been growing interest in the effects of a finite lifetime of particles on statistical properties of random walk and of its continuous limit, the Brownian motion \cite{Lindenberg2013a,Lindenberg2013b,M2015,MR2015,Silverman2016,Grebenkov2017,Silverman2017,Silverman2018}. Practical examples are found in physics, chemistry and biology and vary from diffusion of radioactive gases such as radon \cite{Silverman2016,Silverman2017,Silverman2018} to a variety of search problems such as the search for an oocyte by sperm \cite{MR2015}.  Additional motivation comes from non-equilibrium statistical mechanics: The particle mortality breaks detailed balance and even makes any steady state impossible without a particle source. Last but not least, mortal Brownian motion has long been a classical area of study for mathematicians \cite{BorodinSalminen}.

Here we are telling three short, and we hope instructive, stories about different statistics of mortal Brownian motion.
Story 1 deals with the position statistics of a mortal Brownian particle (MBP) at the time of its spontaneous death.
Story 2 addresses the statistics of a time-integrated quantity: the area under mortal Brownian motion in one dimension. Finally, in story 3 we study the statistics of the winding angle of a MBP around an impenetrable disk on the plane.

In stories 1 and 2 the probability distributions, that we determine, exhibit integrable singularities at zero values of the position and the area, respectively. In story 3 a singularity at zero winding angle appears only in the limit of very high mortality. In this limit there is also an integrable singularity at a nonzero winding angle. This singularity has a different nature. As we show, it is inherited from a recently uncovered singularity of the short-time large-deviation function of the winding angle for \emph{immortal} Brownian motion \cite{MS2019b}.

\section{Story 1: Position distribution}
\label{position}

This story is quite simple, and it sets the stage. Suppose that a MBP with diffusivity $D$ and the decay rate $\mu$ is released at the origin at $t=0$. At time $t=T$, when the particle spontaneously dies, it arrives at a point $\mathbf{x}=\mathbf{X}$. The Brownian motion and the death are independent random processes.
Therefore,
the joint probability distribution $\mathcal{P}(\mathbf{X},T)$ is given by $\mathcal{P}(\mathbf{X},T)=P_1(\mathbf{X},T)P_2(T)$. Here $P_1$ is the probability distribution of immortal Brownian motion to arrive at point $\mathbf{X}$ at time $T$:
\begin{equation}\label{distposition}
P_1(\mathbf{X},T) = \frac{1}{(4\pi D T)^{d/2}}\,e^{-\frac{|\mathbf{X}|^2}{4 DT}},
\end{equation}
where $d$ is the dimension of space. In its turn, the probability distribution $P_2(T)$ follows the exponential decay law:
\begin{equation}\label{distdeath}
P_2(T) = \mu e^{-\mu T} .
\end{equation}
As a result, the joint probability is
\begin{equation}\label{jointdistposition}
\mathcal{P}(\mathbf{X},T) = \frac{\mu}{(4\pi D T)^{d/2}}\,e^{-\frac{|\mathbf{X}|^2}{4 DT}-\mu T}.
\end{equation}
%At fixed $\mathbf{X}\neq 0$ the joint probability tends to zero at $T\to 0$ and $T\to \infty$ and has a maximum at $T \sim |\mathbf{X}|/(\mu D)^{1/2}$.
The probability distribution of the lifetime achievement of the MBP in terms of $\mathbf{X}$
is obtained by integrating $\mathcal{P}(\mathbf{X},T)$ over all possible death times:
\begin{equation}\label{dpall}
p(\mathbf{X}) = \int_0^{\infty} \mathcal{P}(\mathbf{X},T)\,dT= \mu\int_0^{\infty} (4\pi D T)^{-d/2}\,e^{-\frac{|\mathbf{X}|^2}{4 DT}-\mu T}\,dT.
\end{equation}
Notice that, when considered as a function of $\mu$,  $p(\mathbf{X})$ is equal to $\mu$ times the Laplace transform of $P_1(\mathbf{X},T)$, where $\mu$ serves as the parameter of the Laplace transform, and the integration is over time.  This is a generic feature of a whole class of MBP statistics, including those considered in our three stories.

Evaluating the integral in Eq.~(\ref{dpall}), we obtain
\begin{numcases}
{\!\! p(\mathbf{X})=}\sqrt{\frac{\mu}{4D}}\,e^{-\sqrt{\frac{\mu}{D}}|\mathbf{X}|},
%\frac{\sin^2 \theta}{w^2},
& $d=1$, \label{1d}\\
\frac{\mu}{2\pi D} K_0\left(\sqrt{\frac{\mu}{D}}|\mathbf{X}|\right),   & $d=2$, \label{2d} \\
\frac{\mu}{4\pi D |\mathbf{X}|}\,e^{-\sqrt{\frac{\mu}{D}}|\mathbf{X}|},  &
$d=3$, \label{3d}
\end{numcases}
where $K_0(\dots)$ is the modified Bessel function of the second kind. In all dimensions $p(\mathbf{X})$ is isotropic, and it decays exponentially at large $|\mathbf{X}|$. The distribution $p(\mathbf{X})$ exhibits the scaling behavior
\begin{equation}\label{scalingX}
p(\mathbf{X})= \frac{1}{\ell^d} \,\Phi_d \left(\frac{|\mathbf{X}|}{\ell}\right).
\end{equation}
The intrinsic time scale $T\sim 1/\mu$ and the diffusivity $D$ of mortal Brownian motion determine the characteristic decay length $\ell=\sqrt{D/\mu}$. As expected on the physical grounds, a higher mortality leads to a  stronger localization of the distribution near the origin.

Noticeable in Eqs.~(\ref{1d})-(\ref{3d}) is a singularity of $p(\mathbf{X})$ at $|\mathbf{X}|=0$. In one dimension $p(\mathbf{X})$ is bounded at $|\mathbf{X}|=0$, but its first derivative with respect to $\mathbf{X}$ is discontinuous there. In two dimensions  $p(\mathbf{X})$ diverges logarithmically at $|\mathbf{X}|=0$, while in three dimensions it diverges as $1/|\mathbf{X}|$. These singularities, however, are integrable.

Interestingly, the same results (\ref{1d})-(\ref{3d}) are given by the normalizable solutions of the \emph{stationary} diffusion-decay equation with a $d$-dimensional delta-function particle source at $\mathbf{X}=0$:
\begin{equation}\label{diffdecayeq}
\nabla^2 p(\mathbf{X}) -\frac{\mu}{D} p(\mathbf{X})=-\frac{\mu}{D} \delta(\mathbf{X}),
\end{equation}
The normalizable solutions automatically obey the normalization condition $\int p(\mathbf{X}) \,d\mathbf{X} =1$. Equation~(\ref{diffdecayeq}) is apparently valid for any $d$.

\section{Story 2: Area distribution}
\label{area}

\begin{figure}[ht]
\includegraphics[width=0.4\textwidth,clip=]{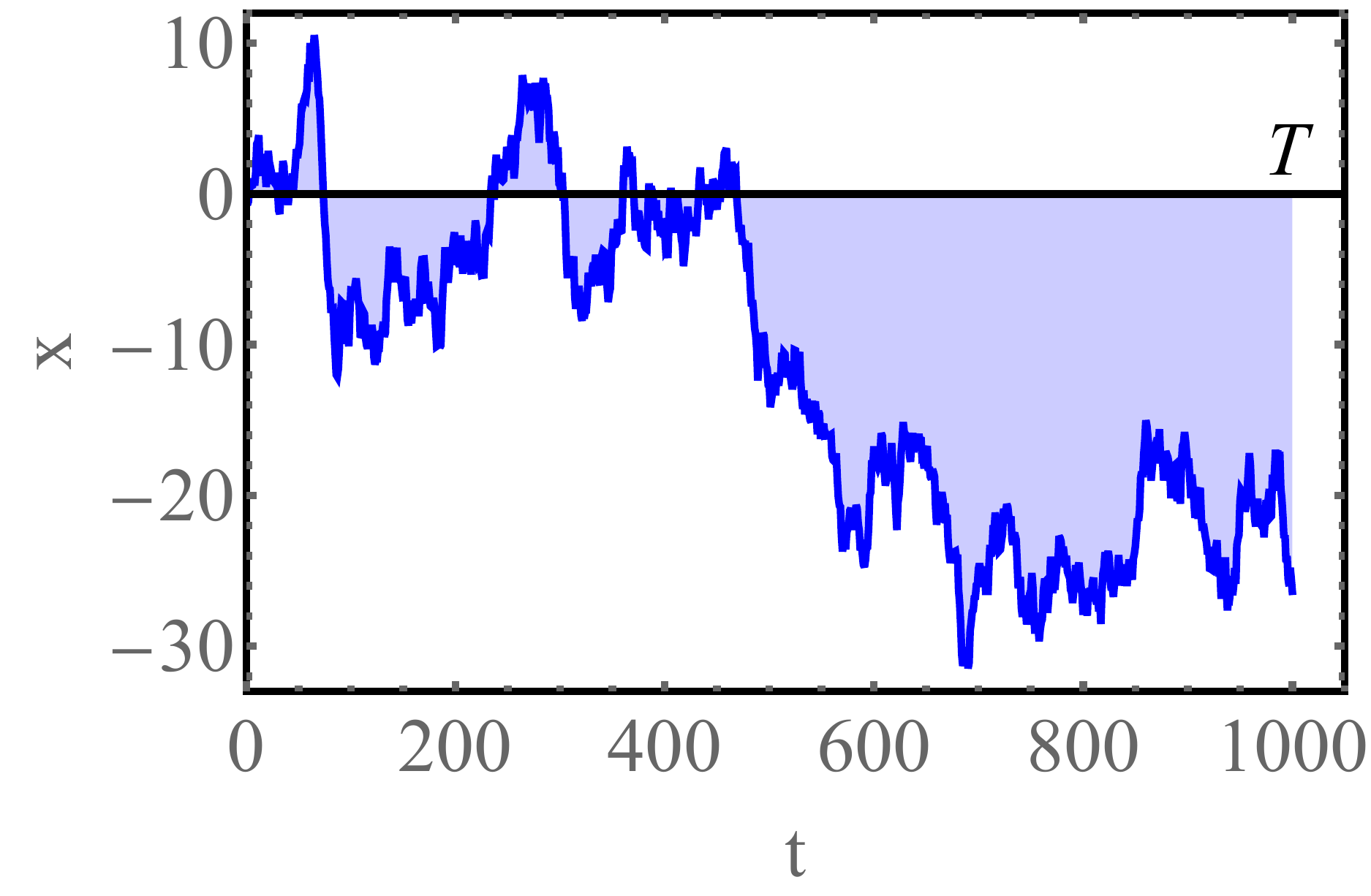}
\caption{A realization of mortal Brownian motion with diffusivity $D=1$. The particle died at $t=T=10^3$. We study the statistics of the area (\ref{area}) of the shaded region. Areas under the $t$-axis are negative.}
\label{BMarea}
\end{figure}

Here a MBP is released at the origin at $t=0$ and allowed to move on the line $-\infty<x<\infty$. This time we are interested in the statistics of a time-integrated quantity: the area
\begin{equation}\label{area}
A=\int_0^T x(t) dt
\end{equation}
that the particle accumulates by the time of its spontaneous death at $t=T$, see Fig. \ref{BMarea}. The joint probability distribution $\mathcal{P}(A,T)=P_1(A,T)P_2(T)$ is again a product of two distributions.  $P_1(A,T)$ is the probability distribution of immortal Brownian motion accumulating the area $A$ by time $T$, whereas $P_2(T)$ is the probability distribution of the particle death at $t=T$, given by Eq.~(\ref{distdeath}).  Now,  $x(t)$ is a Gaussian random variable with zero mean, and the same is true for $A=A(T)$ from Eq.~(\ref{area}). Therefore, the distribution $P_1(A,T)$ can be calculated in a straightforward way by computing the second moment $\langle A^2\rangle$ (see the Appendix), and we obtain
\begin{equation}\label{distarea0}
P_1(A,T) = \frac{\sqrt{3}}{\sqrt{4 \pi DT^3}} e^{-\frac{3 A^2}{4D T^3}}.
\end{equation}
$P_1(A,T)$ is a simple member of a whole family of area distributions of immortal Brownian motion. Other members of this family (the Brownian bridge, the Brownian excursion, the positive part of the Brownian motion, the absolute value of the Brownian motion, \textit{etc}.) are subject to additional constraints, see Ref.  \cite{Janson2007} and references therein. All these distributions exhibit the same scaling behavior as the simple Brownian motion that we are dealing with:
\begin{equation}\label{P1(A)}
P_1(A,T)=\frac{1}{\sqrt{DT^3}} f\left(\frac{A}{\sqrt{D T^3}}\right),
\end{equation}
as to be expected from dimensional analysis and scale invariance of immortal Brownian motion.

\begin{figure}[ht]
\includegraphics[width=0.4\textwidth,clip=]{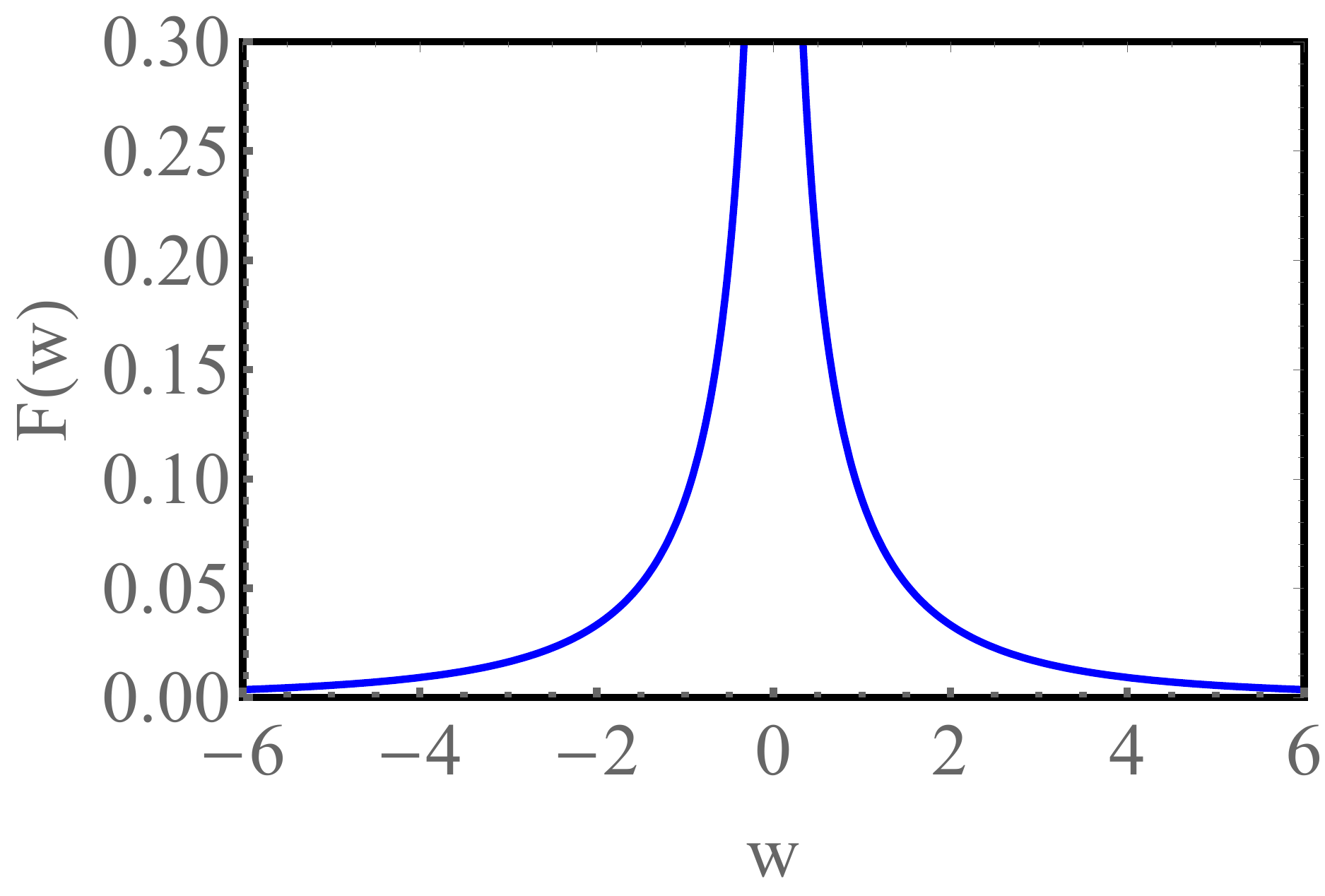}
\caption{The scaling function $F(w)$ of the area distribution $p(A)$ of mortal Brownian motion. See Eqs.~(\ref{p(A)scaling}) and (\ref{F(w)}) for definitions.}
\label{scalingfunction}
\end{figure}

Using Eqs.~(\ref{distdeath}) and (\ref{distarea0}), we obtain the joint probability
\begin{equation}\label{jointdistarea}
\mathcal{P}(A,T) = \frac{\sqrt{3}\mu}{\sqrt{4 \pi DT^3}}\,e^{-\frac{3 A^2}{4D T^3}-\mu T}.
\end{equation}
%At fixed $A\neq 0$ the joint probability vanishes at $T\to 0$ and $T\to \infty$ and has a maximum at $T \sim |A|^{1/2}/(\mu D)^{1/4}$.
Now we turn to the probability distribution $p(A)$ of the lifetime achievement of the MBP. The presence of the intrinsic time scale $T\sim \mu^{-1}$ of mortal Brownian motion implies the following scaling behavior of $p(A)$:
\begin{equation}\label{p(A)scaling}
p(A) = \frac{1}{A_{\mu}}\,F\left(\frac{A}{A_{\mu}}\right),\quad \text{where} \quad A_{\mu}=D^{1/2} \mu^{-3/2}.
\end{equation}
This is indeed what the calculation shows. The distribution $p(A)$ is given by the integral over death times
\begin{equation}\label{distpositionall}
p(A) = \int_0^{\infty} \mathcal{P}(A,T)\,dT= \int_0^{\infty} \frac{\sqrt{3}\mu}{\sqrt{4 \pi DT^3}}\,e^{-\frac{3 A^2}{4D T^3}-\mu T}\,dT,
\end{equation}
which can be evaluated exactly. The result can be written as Eq.~(\ref{p(A)scaling}), where
%\begin{equation}\label{distarea}
%p(A)=\frac{1}{|A|}\,F\left(\frac{3\mu^{3/2}|A|}{D^{1/2}}\right),
%\end{equation}
%where
the scaling function is
\begin{equation}\label{F(w)}
F(w)=\frac{2\pi  \cdot 3^{1/6} \text{Ai} \left(e^{i \frac{\pi}{6}} |3 w|^{1/3}\right) \text{Ai} \left(e^{-i \frac{\pi}{6}}|3 w|^{1/3}\right)}{|w|^{1/3}},
\end{equation}
%\begin{equation}\label{F(w)}
%F(w)=\frac{2\pi  \cdot 3^{1/6} \text{Ai} \left( e^{i \pi/6} |3 w|^{1/3}\right) \text{Ai} \left(e^{-i \pi/6}|3 w|^{1/3}\right)}{|w|^{1/3}},
%\end{equation}
and $\text{Ai}(\dots)$ is the Airy function.  Note that the product $\text{Ai}(z) \text{Ai}(z^*)$, where $z$ is any complex number, is a real number. The function $F(w)$ has the following asymptotics:
\begin{numcases}
{\!\! F(w)\simeq}\frac{2\pi}{3^{7/6} [\Gamma(2/3)]^2 |w|^{1/3}},
& $|w|\ll 1$, \label{wsmall}\\
\frac{1}{2\sqrt{|w|}}\,e^{-2\sqrt{\frac{2 |w|}{3}}},   & $|w|\gg 1$. \label{wlarge}
\end{numcases}
Interestingly, the scaling function $F(w)$ and, as a result, the distribution $p(A)$, diverges at $w=0$ (correspondingly, at $A=0$), see Eq.~(\ref{wsmall}). This singularity is integrable, so that $\int_{-\infty}^{\infty} F(w) dw =1$ as it should be. The asymptotic (\ref{wsmall}) can be obtained directly from Eq.~(\ref{distpositionall}), by neglecting $\mu T\ll 1$ in the exponent. The stretched exponential tail (\ref{wlarge}),  including the pre-exponential factor, can be obtained by evaluating the integral over $T$ in Eq.~(\ref{distpositionall})  by the Laplace method.
Figure \ref{scalingfunction} shows the scaling function $F(w)$.

What can be said about the lifetime achievements in terms of the area distribution of other variants of mortal Brownian motion? Here we will briefly consider the \emph{absolute value} $|x(t)|$ of mortal Brownian motion on the line. To start with, it has the same scaling form (\ref{p(A)scaling}) as that of mortal Brownian motion itself. Furthermore, the large-$A$ tails of the area distributions of \emph{immortal} Brownian motion and of its absolute value on any finite time interval $0<t<T$ coincide: both behave as $\exp[-3A^2/(4DT^3)]$ up to pre-exponential factors. As a consequence, the Laplace method leads to identical results, up to pre-exponential factors,  for the stretched exponential tails $|A|/A_{\mu}\gg 1$ in the mortal versions of the two problems. Both tails behave as %$e^{-2\sqrt{\frac{2 |w|}{3}}}$.
$\sim \exp(-2\sqrt{2|w|/3})$, where $w=A/A_{\mu}=D^{-1/2}\mu^{3/2} A$.

The coincidence of the tails in the immortal case is explained by the fact that, for very large $|A|$, the probability of observing a given $|A|$ is dominated by a single Brownian trajectory $x(t)$. This \emph{optimal} trajectory can be found, in the spirit of geometrical optics, by minimizing the Wiener's action (see, \textit{e.g.} Ref. \cite{legacy})
\begin{equation}\label{action}
S = -\frac{1}{4D} \int_0^T \dot{x}^2 dt
\end{equation}
subject to the initial condition $x(0)=0$ and the following constraints: $\int_0^T x(t) dt =A$ for the Brownian motion, and $\int_0^T |x(t)| dt =A$ for its absolute value. The constraints can be accommodated via Lagrangian multipliers, and the ``lacking" boundary condition at $t=T$ becomes, in  both cases, $\dot{x} (T)=0$ \cite{Elsgolts}. For the Brownian motion the optimal trajectory, for any sign of $A$, is
\begin{equation}\label{optimal}
x(t) = \frac{3 A}{2T^3}\,t \left(2T-t\right),\quad 0\leq t\leq T.
\end{equation}
For the absolute value of the Brownian motion  the optimal trajectory is exactly the same, except that $A$ must be positive. The action (\ref{action}) along this trajectory is $S = 3A^2/(4DT^3)$. It leads to identical distribution tails $P(|A|\gg \sqrt{D T^3})\sim\exp[-3A^2/(4DT^3)]$ (up to pre-exponential factors) in the two problems. We refer the reader to Refs. \cite{GF,SmithMeerson2019,Meerson2019,MS2019b} for other applications of geometrical optics of constrained immortal Brownian motion.

\section{Story 3: Winding angle distribution}
\label{winding}

\begin{figure}[ht]
\includegraphics[width=0.4\textwidth,clip=]{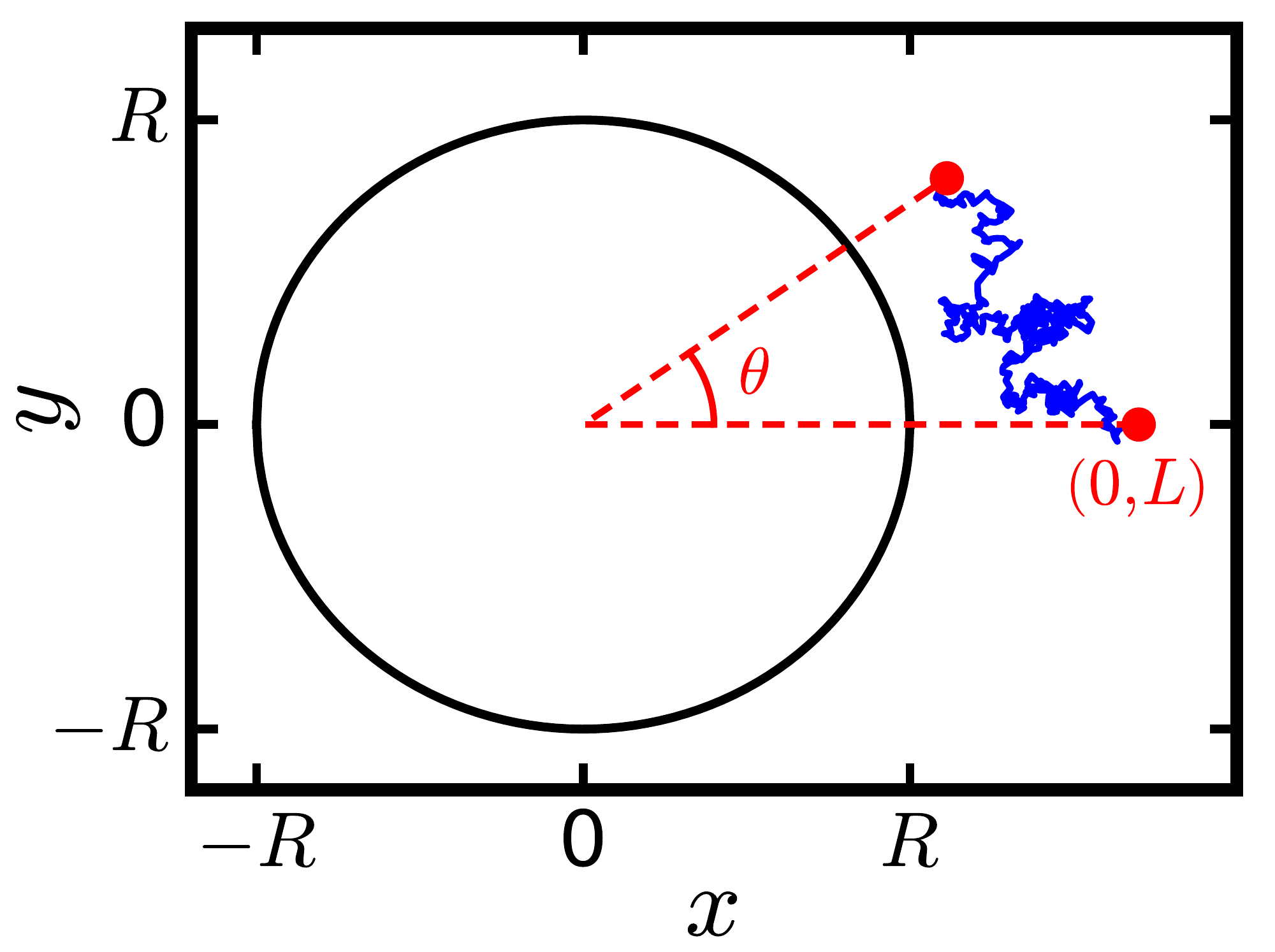}
\caption{A realization of mortal Brownian motion around a reflecting disk with radius $R$.  We study the distribution of the winding angle $\theta$ accumulated by the time of particle death.}
\label{windingangle}
\end{figure}

Suppose that a MBP is released at $t=0$ at a distance $L$ from the center of a reflecting disk with radius $R<L$ in the plane. %One can
%always choose the polar angle of the release point to be zero.
We are interested in the probability distribution $p(\theta)$ of the winding angle $\theta$ of the particle around the disk at the time of particle death, see Fig.~\ref{windingangle}.

The joint probability of the winding angle $\theta$ and of the death time $T$ is $\mathcal{P}(\theta,T)=P_1(\theta,T)P_2(T)$, where $P_1\left(\theta,T\right)$  is the probability distribution  of reaching the winding angle $\theta$ at time $T$ for immortal Brownian motion, and $P_1(T)$ is given by Eq.~(\ref{distdeath}).  The distribution $P_1\left(\theta,T\right)$  has been studied in many works \cite{Rudnick,Belisle,Saleur,GF,Kundu,MS2019b}. In principle, it can be determined exactly by solving the diffusion equation subject to the reflecting boundary condition on the disk and a delta-function initial condition. However, the resulting exact expressions for $P_1\left(\theta,T\right)$, obtained in Refs.~\cite{Rudnick,GF}, include triple integrals of combinations of Bessel functions and trigonometric and/or exponential functions, and they are too cumbersome for our purposes.  Here we will confine ourselves to evaluating  the joint distribution $\mathcal{P}(\theta,T)$ and the achieved winding angle distribution $p(\theta)$ in the limits of
very low and very high mortality, where we can rely on the previously obtained long- and short-time asymptotics of $P_1(\theta,T)$.

The long-time asymptotic of $P_1\left(\theta,T\right)$ corresponds to the strong inequalities  $(DT)^{1/2}\gg R, L$. In this limit  $P_1\left(\theta,T\right)$ does not depend on $L$ \cite{Rudnick,Belisle,Saleur,GF,Kundu}:
\begin{equation}\label{longtime}
P_1\left(\theta,T\right)\simeq\frac{\pi}{2 \ln \frac{4DT}{R^2}}\,\text{sech}^{2}\left(\frac{\pi \theta}{\ln \frac{4DT}{R^2}}\right), \quad \sqrt{DT}\gg R, L.
\end{equation}

The short-time asymptotic of $P_1(\theta,T)$  holds when $(DT)^{1/2} \ll R, L-R$. It has been determined recently in the geometrical-optics approximation \cite{MS2019b}. At such short times a sizable winding angle is a large deviation, and the probability distribution $P_1(\theta,T)$ can be written, up to a pre-exponential factor, as
\begin{equation}\label{shorttime}
-\ln P_1 (\theta, T) \simeq \frac{R^2}{4 DT}\,g\left(\theta, \frac{R}{L}\right),\quad \sqrt{DT}\ll R, L-R,
\end{equation}
where
\begin{numcases}
{\!\! g(\theta, w)=} w^{-2} \sin^2 \theta,
%\frac{\sin^2 \theta}{w^2},
& $\left|\theta\right|\leq\arccos w$, \label{less2}\\
\left(\left|\theta\right|+\sqrt{1/w^2-1}-\arccos w\right)^2,   & $\left|\theta\right|\geq\arccos w$, \label{more2}
\end{numcases}
and $0<w<1$. The rate function $g\left(\theta,w\right)$ and its first $\theta$-derivative are continuous functions, but the second derivative $\partial^2_{\theta} g$ is discontinuous at $\theta=\theta_{\text{c}}=\arccos w$. This singularity of the rate function (which appears only in the limit of $T\to 0$, and is smoothed at finite $T$) can be interpreted as a second-order dynamical phase transition \cite{MS2019b}.

Using Eqs.~(\ref{distdeath}), (\ref{longtime}) and (\ref{shorttime}), we obtain the long- and short-time asymptotics, respectively, of the joint distribution $\mathcal{P}(\theta,T)$.  Let us introduce a dimensionless parameter
$\epsilon = \mu R^2/(4D)$. At very low mortality, $\epsilon\ll 1$, the dominating contribution to the integral
$p(\theta) = \int_0^{\infty} \mathcal{P}(\theta,T)\,dT$ comes from long times, where we can use the asymptotic (\ref{longtime}) of $P_1(\theta,T)$. Upon a change of variables $y=\mu T$ in the integral, we obtain
\begin{equation}\label{pthetaint}
p(\theta) \simeq \frac{\pi}{2} \int_0^{\infty} \frac{dy}{\ln \frac{y}{\epsilon}}\, \text{sech}^{2} \left(\frac{\pi \theta}{\ln \frac{y}{\epsilon}}\right) e^{-y}.
\end{equation}
The characteristic integration length, $O(1)$, of this integral is determined by the exponential $e^{-y}$. On this scale the function $\ln (y/\epsilon)$ changes very little. Therefore,  we can evaluate it at any $y=O(1)$ and obtain, with logarithmic accuracy,
\begin{equation}\label{ptheta1}
 p(\theta) \simeq \frac{\pi}{2 \ln \frac{1}{\epsilon}}\, \text{sech}^{2} \left(\frac{\pi \theta}{\ln \frac{1}{\epsilon}}\right),\quad \epsilon=\frac{\mu R^2}{4D}\ll 1.
\end{equation}
As one can see, the $\mu$-dependence is only logarithmic.

The situation is very different at high mortality, $\epsilon\gg 1$. Using Eq.~(\ref{shorttime}) we obtain, up to a pre-exponential factor,
\begin{equation}\label{ptheta2}
p(\theta) \sim \int_0^{\infty} e^{-\frac{R^2}{4 DT}g\left(\theta, \frac{R}{L}\right)-\mu T} dT
\end{equation}
Evaluating the integral by the Laplace method, we arrive at
\begin{equation}\label{ptheta3}
p(\theta) \sim \exp\left[-2\sqrt{\epsilon}\, h\left(\theta, \frac{R}{L}\right)\right], \quad \epsilon=\frac{\mu R^2}{4D}\gg 1,
\end{equation}
%
%$p(\theta) \sim \exp[-\sqrt{2\epsilon}\, h(\theta, R/L)]$,
where
\begin{numcases}
{\!\! h\left(\theta, \frac{R}{L}\right)\equiv \sqrt{g\left(\theta,\frac{R}{L}\right)}=} \frac{L |\sin \theta|}{R},
%\frac{\sin^2 \theta}{w^2},
& $\left|\theta\right|\leq\arccos \frac{R}{L}$, \label{less3}\\
\left|\theta\right|+\sqrt{\frac{L^2}{R^2}-1}-\arccos \frac{R}{L},   & $\left|\theta\right|\geq\arccos  \frac{R}{L}$. \label{more3}
\end{numcases}

\begin{figure}[ht]
\includegraphics[width=0.35\textwidth,clip=]{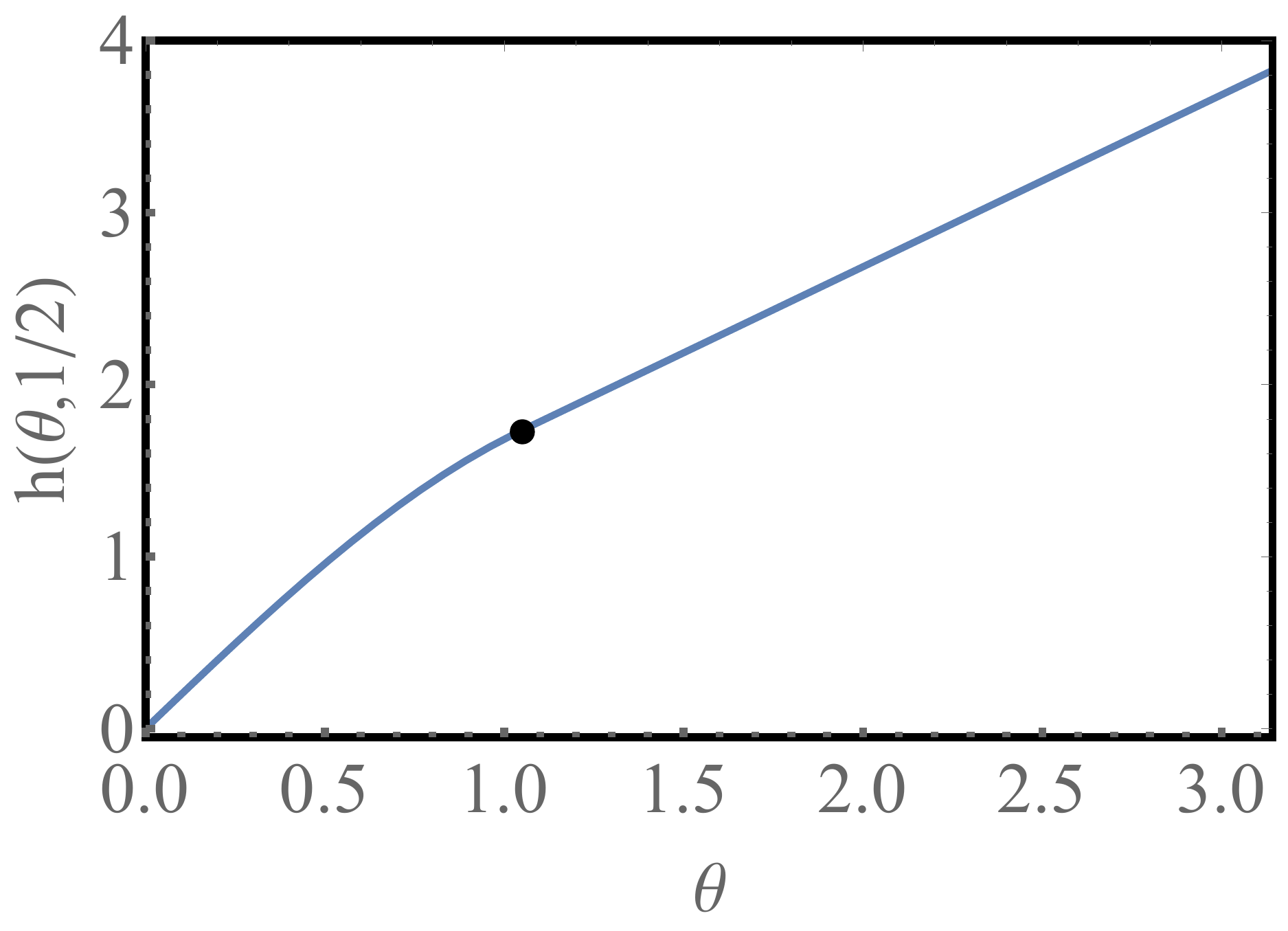}
\caption{The rate function $h(\theta,R/L=1/2)$ of the winding angle distribution $p(\theta)$ of mortal Brownian motion as a function of $\theta$. Only nonnegative $\theta$ are shown. See Eqs.~(\ref{ptheta3})-(\ref{more3}) for definitions. There is a second-order dynamical phase transition at $\theta=\arccos(1/2)$, marked by the fat point.}
\label{htheta}
\end{figure}

Note that the high-mortality rate function  $h(\theta,R/L)$ is everywhere non-convex, see Fig. \ref{htheta}. This is a relatively rare situation. In addition to the corner singularity at $\theta=0$, $h(\theta,R/L)$ exhibits a discontinuity in the second derivative $\partial_{\theta}^2 h$ at $\theta=\arccos (R/L)$ which is inherited from
the singularity of the short-time large-deviation function (\ref{shorttime}). The tail of $p(\theta)$ is exponential. Finally, the $\mu$ dependence here is much stronger than in the low-mortality limit.

\section{Discussion}

\label{disc}

The model of mortal Brownian motion is relevant for a number of practical issues, including diffusion of radioactive materials \cite{Silverman2016,Silverman2017,Silverman2018}. But it is also interesting in a more general context.
That a finite lifetime puts a natural upper limit on the achievements of any individual or a collective of individuals  is almost a truism, and a convenient mathematical framework for this general phenomenon can be useful.  Here we have considered three very basic settings -- which we called three short stories -- dealing with different statistics of a mortal Brownian particle (MBP). The ``lifetime achievements" of the MBP -- in terms of (1) the final position, (2) the accumulated area, and (3) the final winding angle around an obstacle -- are totally determined by chance. As we observed, the corresponding probability distributions are nontrivial.  In stories 1 and 2 they exhibit singularities at zero values of the position and the area, respectively. In story 3 we observed a singularity at zero winding angle only in the limit of very high mortality. In this limit there is also a singularity of a different type. It appears at a nonzero winding angle and results from the singularity of the short-time large-deviation function of the winding angle for immortal Brownian motion \cite{MS2019b}. It would be interesting to extend some of our analysis to \emph{ensembles} of MBPs and, more generally, to ensembles of \emph{interacting} diffusing particles which can be modelled as lattice gases \cite{M2015}. The macroscopic fluctuation theory (see Ref. \cite{MFTreview} for a general review and Ref. \cite{M2015} for some examples) provides an efficient and versatile starting point for such extensions.

\section*{ACKNOWLEDGMENTS}

I am grateful to Tal Agranov and Naftali R. Smith for useful discussions, and to Naftali R. Smith for help with Fig. \ref{windingangle}. This research was supported by the Israel Science Foundation (grant No. 807/16).

\begin{appendices}
\appendix
\section{Derivation of Eq.~(\ref{distarea0})}
\label{appendix:OFMequations}
\renewcommand{\theequation}{A\arabic{equation}}
\setcounter{equation}{0}

We start with the stochastic PDE for the immortal Brownian motion, see \textit{e.g.} Ref. \cite{legacy}:
\begin{equation}\label{Langevin}
\dot{x}(t) = \sqrt{2D} \xi(t),
\end{equation}
where $\xi(t)$ is a Gaussian white noise with zero mean and
\begin{equation}\label{corr}
\langle \xi(t) \xi(t')\rangle =\delta(t-t').
\end{equation}
The solution of Eq.~(\ref{Langevin}) with initial condition $x(0)=0$ is
\begin{equation}\label{x(t)}
x(t) = \sqrt{2D} \int_0^t \xi(t') dt'.
\end{equation}
We are interested in the probability distribution of
\begin{equation}\label{A(t)}
A = \int_0^T x(t) dt = \sqrt{2D} \int_0^T dt \int_0^t dt' \xi(t').
\end{equation}
This distribution is Gaussian, therefore it suffices to compute the variance: $\text{Var} = \langle A^2\rangle-\langle A \rangle$. Since  $\langle A \rangle=0$, we have
\begin{eqnarray}
  \text{Var} &=& \langle A^2\rangle =
  2D \,\Big\langle\int_0^T dt_1 \int_0^T dt_2\int_0^{t_1} dt' \xi(t')\int_0^{t_2} dt'' \xi(t'')\Big \rangle \nonumber\\
  &=& 2D \int_0^T dt_1 \int_0^T dt_2\int_0^{t_1} dt' \int_0^{t_2} dt''  \big \langle\xi(t')\xi(t'') \big \rangle .
\end{eqnarray}
Using Eq.~(\ref{corr}), we obtain
\begin{equation}
  \text{Var} = 2D \int_0^T dt_1 \int_0^T dt_2 \min(t_1,t_2)
  =\frac{2DT^3}{3},
\end{equation}
leading to Eq.~(\ref{distarea0}).

\end{appendices}

\end{document}